\documentclass[preprint,12pt]{elsarticle}

\usepackage{epsfig}
\usepackage{enumerate}
\usepackage[normalem]{ulem}
\usepackage{amssymb,amsmath,graphics,graphicx}  
\usepackage{xcolor}
\usepackage{mathtools}

\usepackage{booktabs} 
\usepackage{multirow}

\journal{Elsevier}

\begin{document}

\begin{frontmatter}

%% Title, authors and addresses

%\title{On the propagation of racism on social media}

%\title{Modeling the dynamics of racism spreading in online social networks}

\title{A Statistical Physics perspective on fairness in shared expenses: The bar bill analogy}

%% use the tnoteref command within \title for footnotes;
%% use the tnotetext command for the associated footnote;
%% use the fnref command within \author or \address for footnotes;
%% use the fntext command for the associated footnote;
%% use the corref command within \author for corresponding author footnotes;
%% use the cortext command for the associated footnote;
%% use the ead command for the email address,
%% and the form \ead[url] for the home page:
%%
%% \title{Title\tnoteref{label1}}
%% \tnotetext[label1]{}
%% \author{Name\corref{cor1}\fnref{label2}}
%% \ead{email address}
%% \ead[url]{home page}
%% \fntext[label2]{}
%% \cortext[cor1]{}
%% \address{Address\fnref{label3}}
%% \fntext[label3]{}

%% use optional labels to link authors explicitly to addresses:
%% \author[label1,label2]{<author name>}
%% \address[label1]{<address>}
%% \address[label2]{<address>}

\author{Nuno Crokidakis}
\ead{nunocrokidakis@id.uff.br}
\author{Lucas Sigaud}

\address{
Instituto de F\'{\i}sica, Universidade Federal Fluminense, Niter\'oi, Rio de Janeiro, Brazil
}

\begin{abstract}
%% Text of abstract
In social contexts where individuals consume varying amounts, such as shared meals or bar gatherings, splitting the total bill equally often yields surprisingly fair outcomes. In this work, we develop a statistical physics framework to explain this emergent fairness by modeling individual consumption as stochastic variables drawn from a realistic distribution, specifically the gamma distribution. Introducing a Boltzmann-like weighting factor, we derive exact analytical expressions for the partition function, average consumption, variance, and entropy under economic or social penalization constraints. Numerical simulations, performed using the Marsaglia-Tsang algorithm, confirm the analytical results with high precision. Drawing a direct parallel between individual consumption and ideal gas particle energy in the canonical ensemble, we show how the law of large numbers, mutual compensation, and the effective ordering induced by penalization combine to make equal cost-sharing statistically robust and predictable. These findings reveal that what appears to be an informal social convention is, in fact, grounded in the same fundamental principles that govern the collective behavior of particles in thermodynamic systems, highlighting the interdisciplinary power of statistical physics.

\end{abstract}

\begin{keyword}
Dynamics of social systems \sep Collective phenomena \sep Statistical Physics \sep Monte Carlo simulations \sep Canonical ensemble
%% keywords here, in the form: keyword \sep keyword

%% MSC codes here, in the form: \MSC code \sep code
%% or \MSC[2008] code \sep code (2000 is the default)

\end{keyword}

\end{frontmatter}

%%
%% Start line numbering here if you want
%%
%\linenumbers

%% main text
\section{Introduction}

\quad Decision-making in groups and resources allocation are fundamental phenomena in both economic and social systems, in large and small scales, and many physical statistical models are employed in their description~\cite{barthe}. An everyday familiar context in which both group decisions and resource allocation are commonly employed is in the division of a shared bill in dining scenarios, such as when friends dine together at a bar or restaurant. Although individuals usually consume different amounts, ordering more or less food and drinks than others, with varying prices, it is a common practice in many places in the world to split the total cost equally among the individuals. Perhaps surprisingly, this approach is often perceived as a satisfactory, or even fair, way to pay for the bill by most participants. A question that must be posed is: why does such a simple and apparently naive rule often work so well?

%\quad Group decision-making and resource allocation are fundamental phenomena in both economic and social systems. One particularly familiar context is the division of a shared bill in group dining scenarios, such as when friends dine at a bar or restaurant. Although individuals may consume varying amounts, some ordering more food or drinks than others, it is common practice to split the total cost equally. Surprisingly, this approach is often perceived as fair or at least satisfactory by most participants. Why should such a simple and seemingly naive rule work so well?

Although the question of bill-splitting has been previously addressed in the context of behavioral economics, particularly in the study by Gneezy, Haruvy, and Yafe \cite{gneezy}, the focus has largely been on the inefficiencies and incentive distortions associated with equal division. In their experimental study, the authors showed that individuals tend to consume more when the cost is shared equally, compared to when each person pays for their own consumption. While this line of research sheds light on behavioral responses to shared expenses, it does not address the deeper statistical structure underlying the relative fairness of equal sharing in spontaneous, uncoordinated group settings. In contrast, the present work aims to explore this question through the lens of statistical physics, highlighting how collective regularities and fairness can emerge even in the absence of explicit coordination.

Therefore, we will here endeavor to study the deeper statistical regularities that emerge in collective behavior. This relates directly to thermodynamic systems described by statistical mechanics, where individual-level randomness gives rise to predictable macroscopic outcomes, even in the presence of outliers if the sample is large enough. Statistical physics models provide a robust framework for understanding the macroscopic regularities that emerge from microscopic disorder. In the canonical ensemble, the energy of each particle fluctuates due to thermal contact with a reservoir, but the mean energy per particle becomes well defined as the number of particle increases~\cite{huang,pathria}. This averaging mechanism, driven by the law of large numbers, has many analogues in social and economic systems, where individual decisions may have large variability but group behaviors display stable and predictable outcomes~\cite{huang,pathria,rmp,pmco,galam_review,galam_book,sen_book}.

%This question touches upon deeper statistical regularities that emerge in collective behavior. As in thermodynamic systems studied in statistical mechanics, individual-level randomness can give rise to predictable macroscopic outcomes. Statistical physics provides a robust framework for understanding how macroscopic regularity emerges from microscopic variability. In the canonical ensemble, the energy of each particle fluctuates due to thermal contact with a reservoir, but the mean energy per particle becomes sharply defined as the number of particles increases \cite{huang,pathria}. This averaging mechanism, driven by the law of large numbers, has analogues in social systems, where individual decisions may vary but group-level outcomes display stability and predictability \cite{huang,pathria,rmp}.

Recent developments in sociophysics and econophysics have extended these ideas to contexts far beyond the physical sciences. Models of social interaction, opinion dynamics, and economic exchange often rely on similar stochastic foundations \cite{rmp,helbing,gallegati}. The thermodynamic foundation of social laws is based on the law of statistics under constraints, which in physics it is called the "free energy principle"~\cite{mimkes}. In social sciences it may be called the "principle of maximum happiness of societies", and may be applied to collective or individual behavior of social, political or religious groups. For instance, Helbing \cite{helbing} and Castellano et al. \cite{rmp} describe how individual-level interactions governed by simple probabilistic rules can produce emergent order in social systems. In the context of shared expenses, a similar mechanism may be at work: despite variability in individual consumption, aggregate behavior conforms to regular patterns that make equal division seem reasonable. Moreover, studies in complex networks and interacting particle systems \cite{barrat} show that even weak correlations or interactions among units can shape collective dynamics significantly \cite{bejan}. %In the dining scenario, social influences, such as mutual awareness, fairness norms, and menu constraints, create mild correlations between individuals' choices. These correlations can be understood as analogous to weak interactions in physical systems (e.g., van der Waals gases), which modify but do not eliminate ensemble properties.

In dining scenarios, many social influences drive the group behavior as they create correlations, even if mild ones, between individuals' choices. Among such social influences, we can list mutual awareness, fairness norms, menu constraints, guilt and an overall pattern on drinking and dining among others~\cite{benjamin}. The correlations created can be compared to the weak interactions in physical thermodynamical systems (e.g. van der Waals gases), which modify and influence the ensemble properties, but without eliminating them. The interesting aspect of choosing such an approach to bar-bill-splitting is that the analogy lies in the connection between the amount due to each individual and the amount actually paid by all individuals in a equally-divided check; but the amount due to each individual is directly linked to particularities in their own behavior, such as drinking beer, whiskey or non-alcoholic drinks, different types of food, dessert ordering pattern, etc.

In this work, we explore this intricacies in the human behavior and how it relates to the overall group pattern that emerge using statistical physics, in particular the thermodynamical model using the canonical ensemble, applied to the equal splitting of a bar or restaurant bill. In particular, we introduce a penalization parameter $\alpha$, which is an analogue to the thermodynamical quantity $\beta$, that corresponds to the social pressure against spending much more than the rest of the individuals. We show that fairness among individuals in equal-splitting bills emerges as a consequence of the statistical averaging commonly used in energy distributions in thermodynamics.

%%%%%%%%%%%%%%%%%%%%%%%%%%%

\section{Statistical model and Thermodynamic analogy}
\quad We model individual consumption as a non-negative random variable $x_i$, drawn from a distribution $\rho(x)$ with mean $\mu$ and variance $\sigma^2$. In a group of $N$ individuals, the total expenditure $E$ and the equal share paid $\langle x\rangle$ by each person are, respectively,
\begin{eqnarray} \label{eq1}
E & = & \sum_{i=1}^{N} x_i \\ \label{eq2}
\langle x\rangle & = & \frac{E}{N} = \frac{1}{N}\,\sum_{i=1}^{N} x_i
\end{eqnarray}

We are interested in how closely $\langle x\rangle$ approximates each individual's actual consumption $x_i$, and what governs the magnitude of the deviation $|x_i - \langle x\rangle|$. This aligns with the canonical ensemble in statistical physics, where each particle's energy $E_i$ is drawn from a Boltzmann-weighted distribution. The total energy $E$ fluctuates, but the average $\langle E \rangle = E/N$ becomes stable. We assume that each individual has a consumption distribution $\rho(x_i)$, and that the psychological or social cost of spending more is penalized by a factor $\alpha$, analogous to the inverse temperature Boltzmann factor $\beta$. Thus, we define a penalized probability for consumption:
\begin{equation} \label{eq3}
P(x_i) = \frac{1}{Z} e^{-\alpha x_i} \rho(x_i),
\end{equation}
\noindent
with $\alpha$ encoding economic or social pressure, analogous to inverse temperature. The partition function is then defined as:
\begin{equation} \label{eq4}
Z_{\alpha} = \int_0^\infty e^{-\alpha x} \rho(x)\,dx.
\end{equation}

This allows derivation of canonical expected values, namely the mean consumption $\langle x\rangle_{\alpha}$ and its variance $\sigma^{2}_{\alpha}=\langle x^{2}\rangle_{\alpha} - \langle x\rangle^{2}_{\alpha}$, as \cite{huang}
\begin{eqnarray} \label{eq5}
\langle x \rangle_{\alpha}  & = & -\frac{\partial}{\partial \alpha} \ln Z_{\alpha} \\ \label{eq6}
\sigma^{2}_{\alpha} & = &  \frac{\partial^{2}}{\partial \alpha^{2}} \ln Z_{\alpha}
\end{eqnarray}

These canonical expressions establish a formal analogy between the bar consumption model and classical statistical physics models. Specifically, the way in which the penalization parameter $\alpha$ influences individual expenditures parallels how the inverse temperature $\beta$ controls the energy distribution in an ideal gas within the canonical ensemble. We explore this correspondence in Tab. \ref{Tab1}. The Boltzmann factor $e^{-\alpha\,x_i}$ in our model penalizes large consumptions, in a manner similar to how the canonical Boltzmann factor $e^{-\beta\,E_i}$ penalizes high-energy microstates in the gas. For low values of $\alpha$ (analogous to high temperature), the penalization is weak: expenditures are unconstrained, leading to wide variability and a highly disordered state. Conversely, high values of $\alpha$ (low temperature) imply strong penalization: expenditures become concentrated at lower values, reflecting an emergent ordering similar to how particles in a gas at low temperature predominantly occupy low-energy states. The conceptual correspondence summarized in Tab. \ref{Tab1} not only clarifies the role of the penalization parameter $\alpha$, but also highlights how consumption homogenization under strong penalization mirrors the shift toward lower energy states observed in thermodynamic systems at low temperatures.

%%%%%%%%%%%%%%%%%%%%%%%%%%%%%%%%%%%%%%%%%%%%%%%%%%%%%%%%%%%%
\begin{table*}[tb]
\centering
\scriptsize
%\footnotesize
\renewcommand\arraystretch{1.05}
\setlength{\tabcolsep}{6pt}
\begin{tabular}{|l|l|l|}
\hline
\textbf{Concept} & \textbf{Bar consumption model} & \textbf{Ideal gas model (canonical)} \\ \hline
Inverse temperature & Penalization parameter $\alpha$ & $\beta = 1/T$ (inverse temperature) \\
Variable & Individual consumption $x_i$ & Particle energy $E_i$ \\
Energy interpretation & Consumption as energy cost & Kinetic energy of particles \\
Boltzmann factor & $e^{-\alpha x_i}$ penalizes high consumption & $e^{-\beta E_i}$ penalizes high-energy states \\
Partition function & $Z_{\alpha} = \int e^{-\alpha x} \rho(x)\,dx$ & $Z_{\beta} = \int e^{-\beta E} g(E)\,dE$ \\
Low $\alpha$ / High $T$ & Unconstrained, heterogeneous expenditures & Broad energy distribution \\
High $\alpha$ / Low $T$ & Aligned behavior, similar expenditures & Particles concentrate in low-energy states \\
\hline
\end{tabular}
\caption{Analogy between the bar consumption model and the canonical ideal gas, illustrating how penalization of high expenditures parallels thermodynamic constraints on energy at different temperatures.}
\label{Tab1}
\end{table*}
%%%%%%%%%%%%%%%%%%%%%%%%%%%%%%%%%%%%%%%%%%%%%%%%%%%%%%%%%%%%

Having established the general statistical framework, including the definition of the partition function and expressions for statistical averages, we now proceed to apply these results to a specific choice of the individual spending distribution $\rho(x)$. In particular, we choose the gamma distribution, that is a natural and realistic choice for modeling individual expenditures in contexts such as bar or restaurant consumption. First, it is defined strictly on the positive real axis ($x>0$), reflecting the fact that expenditures cannot be negative. Second, it is inherently asymmetric, capturing the typical empirical observation that while most individuals spend near a typical value, there is a long tail of fewer individuals with significantly higher expenditures (e.g., due to ordering premium items). Third, the gamma distribution has two parameters ($k$ and $\theta$) that independently control the typical scale of consumption (mean) and the variability (shape), allowing it to flexibly fit a wide range of empirical data. Finally, its analytical tractability enables closed-form expressions for key thermodynamic quantities such as the partition function, mean, variance, and entropy, facilitating both theoretical analysis and numerical validation within the proposed framework.

%%%%%%%%%%%%%%%%%%%%%%%%%%%

\section{The maximum relative entropy method: A natural way of explanation}

\quad In the previous section we defined our ``penalized'' probability for consumption in the form given by Eq. \eqref{eq3}, with the corresponding partition function given by Eq. \eqref{eq4}.

However, such two equations can be obtained from the maximum relative entropy method, which make the motivation for the proposed model even clearer. Historically, the maximum relative entropy method is a direct descendant of the MaxEnt method, pioneered by Jaynes \cite{Jaynes1,Jaynes2,Sivia,Caticha}.

We can start from a prior model $\rho(x)$ and update it by maximizing the relative entropy functional \cite{Caticha,Pachter}
\begin{equation} \label{new_eq1}
\mathcal{S}[P,\rho] = - \int_0^\infty P(x)\,\ln\frac{P(x)}{\rho(x)}\,dx
\end{equation}
The probability $P(x)$ is subject to the constraint
\begin{equation} \label{new_eq2}
\langle x \rangle = \int_0^\infty x\,P(x)\,dx = x_0
\end{equation} 
\noindent
where $x_0$ is the the known value of the average individual consumption. Maximizing $\mathcal{S}[P,\rho]$, Eq. \eqref{new_eq1}, over $P(x)$ subject to \eqref{new_eq2} and normalization, yields \cite{Caticha}
\begin{equation} \label{new_eq3}
P(x) = \frac{1}{\int_0^\infty e^{-\lambda x}\,\rho(x)\,dx}\,e^{-\lambda x}\,\rho(x)   
\end{equation}
where $\lambda$ is a Lagrange multiplier. This distribution is normalized, and we have the partition function $Z$ as the normalization factor,
\begin{equation} \label{new_eq4}
Z = \int_0^\infty e^{-\lambda x} \rho(x)\,dx
\end{equation}
Thus, our penalization parameter $\alpha$ emerges as the Lagrange multiplier $\lambda$ imposing the constraint in \eqref{new_eq2}, so its magnitude $|\alpha|$ in fact measures the strength of said constraint, in complete agreement with our initial intuition.

In addition, from the functional \eqref{new_eq1}, one can obtain the entropy $S$,
\begin{equation}  \label{new_eq5}
S = \left\langle - \ln\frac{P(x)}{\rho(x)}\right\rangle    
\end{equation}
\noindent
which leads immediately to $S\leq 0$, with $S=0$ occuring only for $P(x)=\rho(x)$, i.e., for $\alpha=0$ (no penalization). We will discuss the meaning of the negative signal of entropy in the next section.

%%%%%%%%%%%%%%%%%

\section{Canonical analysis of statistical quantities}

\quad We considered a gamma distribution for the probability density $\rho(x)$, i.e., 
\begin{equation} \label{eq7}
\rho(x) = \frac{1}{\Gamma(k)\,\theta^{k}}\,x^{k-1}\,e^{-x/\theta}
\end{equation}
\noindent
In the gamma distribution, the shape parameter $k$ controls the overall form of the distribution. For $k<1$, the distribution is highly skewed with a peak at zero, while for $k>1$ it becomes unimodal with a peak at $(k-1)\,\theta$, approaching a normal distribution as $k$ increases further. The scale parameter $\theta$ stretches or compresses the distribution along the x-axis, directly setting the scale of typical values without altering the relative shape. Together, these parameters allow flexible modeling of positively-defined, right-skewed data such as individual expenditures in social contexts.

The gamma distribution provides a more realistic model for individual expenditures, as it allows for tunable variability through its shape parameter. It encompasses the exponential case as a particular instance ($k=1$) and offers an adjustable shape parameter that governs the degree of individual heterogeneity. Despite its increased complexity compared to the exponential case, it still permits closed-form analytical expressions for the partition function, mean, and relative fluctuations, making it particularly well-suited for analytical and numerical investigations within the canonical framework.

In the context of bar consumption, the parameter $k$ models the degree of homogeneity or heterogeneity among individuals. For small $k$, specifically $k<1$, most individuals spend very little, while a few spend significantly more, leading to a highly skewed distribution. As $k$ increases beyond $k=1$, the expenditures become more evenly distributed around the peak at $(k-1)\,\theta$, and for large $k$, the consumption distribution becomes approximately symmetric and bell-shaped, indicating a more homogeneous group behavior. On the other hand, the scale parameter $\theta$ sets the monetary scale of individual expenditures. Increasing $\theta$ uniformly stretches the distribution, raising both the average consumption ($\langle x\rangle=k\,\theta$) and the standard deviation ($\sigma=\sqrt{k}\,\theta$). In practical terms, it reflects whether individuals are consuming inexpensive items (small $\theta$) or more expensive ones (large $\theta$), without altering the relative shape of the distribution. Together, $k$ and $\theta$ enable flexible modeling of consumption behaviors in group settings, capturing both the typical expenditure scale and the degree of spending inequality within the group. This flexibility makes the gamma distribution a natural choice for studying fairness and collective effects in bar bill division problems.

Following Eq. \eqref{eq4}, the partition function related to the function \eqref{eq7} is given by
\begin{equation} \label{eq8}
Z_{\alpha} = \left(\frac{1}{1+\alpha\,\theta}\right)^{k}
\end{equation}
\noindent
which leads to, from Eqs. \eqref{eq5} and \eqref{eq6},
\begin{eqnarray} \label{eq9}
\langle x\rangle_{\alpha} & = & \frac{k\,\theta}{1+\alpha\,\theta} \\ \label{eq10}
\sigma^{2}_{\alpha} & = & \frac{k\,\theta^{2}}{(1+\alpha\,\theta)^{2}} 
\end{eqnarray}
\noindent
Notice that, for the unpenalized case ($\alpha=0$), we recover the above mentioned results for the gama distribution, namely mean  $\langle x\rangle=k\,\theta$ and standard deviation $\sigma=\sqrt{k}\,\theta$. We also note that both the mean and the variance decrease monotonically with increasing $\alpha$. The canonical mean $\langle x \rangle_{\alpha} = k\theta / (1 + \alpha\theta)$ is inversely proportional to the penalization parameter, indicating that higher $\alpha$ values suppress individual expenditures, consistent with the interpretation of $\alpha$ as a behavioral constraint. This reflects the social or economic pressure embedded in the model, whereby stronger penalization discourages excessive consumption. Similarly, the variance $\sigma^2_{\alpha}$ also diminishes with increasing $\alpha$, signaling that not only do individuals spend less on average, but their consumption levels become more homogeneous under stronger penalization.

Beyond the influence of the penalization parameter $\alpha$, the shape $k$ and scale $\theta$ parameters of the gamma distribution also play a crucial role in determining the statistical properties of individual consumption. As seen from Eqs. \eqref{eq9} and \eqref{eq10}, the canonical mean $\langle x \rangle_{\alpha}$ is directly proportional to the product $k\,\theta$, while the variance $\sigma^{2}_{\alpha}$ is proportional to $k\,\theta^{2}$, both modulated by the penalization term. Increasing $k$, which controls the shape or "peakedness" of the gamma distribution, leads to a proportional increase in both the average consumption and its variance. However, it also reduces the relative fluctuations, as measured by the coefficient of variation $\sigma_{\alpha}/\langle x \rangle_{\alpha} \propto 1/k \,$ \footnote{The coefficient of variation (CV) of a random variable $x$, defined as $CV=\sigma/\langle x \rangle$, is a standard measure of relative dispersion used in statistical analysis \cite{forbes}.}. This implies that, although average consumption grows, the distribution becomes more concentrated around the mean, representing more consistent behavior across individuals. On the other hand, increasing $\theta$, which controls the scale of consumption, leads to larger typical expenditures and broader variability. Therefore, $k$ and $\theta$ jointly determine the baseline heterogeneity of consumption in the absence of penalization, upon which $\alpha$ acts to impose collective constraints.

The entropy can also be calculated analytically. From Eq. \eqref{eq3}, one obtains $P(x)/\rho(x)=(1/Z)\,e^{-\alpha x}$, which leads to $\ln(P(x)/\rho(x))=-\alpha\,x - \ln Z$. Considering such result, the entropy can be obtained from Eq. \eqref{new_eq5}, leading to $S_{\alpha}=\alpha\,\langle x\rangle_{\alpha} + \ln Z_{\alpha}$ \footnote{Although discussed here in the context of the gamma distribution, this expression holds exactly for any underlying distribution $\rho(x)$. In cases where analytical evaluation is not feasible, it can still be computed numerically via sampling of $\rho(x)$ combined with the Boltzmann weighting $e^{-\alpha\,x}$.}. Considering Eqs. \eqref{eq8} and \eqref{eq9}, one obtains
\begin{equation} \label{eq11}
S_{\alpha} = \frac{\alpha\,k\,\theta}{1+\alpha\,\theta} - k\,\ln (1+\alpha\,\theta)
\end{equation}
\noindent
It is worth noting that for the unpenalized case ($\alpha=0$), the entropy $S_{\alpha}$ given by Eq. \eqref{eq11} vanishes. This is consistent with the definition used here, where entropy measures the deviation from the unweighted (pure) distribution, effectively acting as a relative entropy with respect to the original density $\rho(x)$. Thus, $S_{\alpha=0}=0$ reflects the absence of any penalization-induced deformation. In analogy with the ideal gas model, $\alpha=0$ corresponds to the infinite temperature limit $\beta=0$, where particles are completely disordered and the thermodynamic entropy reaches its maximum. However, it is important to note that in our definition, entropy is measured relative to the unweighted distribution, and therefore $S=0$ here corresponds to this maximal disorder baseline, whereas increasing $\alpha$ (analogous to lowering temperature in the ideal gas) reduces entropy by enforcing consumption homogenization, similar to how a gas at low temperature has particles predominantly occupying low-energy states with reduced configurational entropy.

In the next section, we discuss and compare these analytical results with numerical simulations.

% ###################################################

\section{Simulation results}

\quad In this section we consider numerical results for the gamma distribution. Although several candidate distributions may be considered to model individual expenditures, such as the exponential, log-normal, or others, the gamma distribution provides an optimal balance between analytical tractability and empirical realism. It encompasses the exponential case as a particular instance ($k=1$) and offers an adjustable shape parameter that governs the degree of individual heterogeneity. Furthermore, all relevant quantities under the canonical ensemble, including the partition function, mean, variance and entropy, admit closed-form expressions, as we showed in the previous section. These features allow for a transparent comparison between analytical predictions and numerical simulations. For these reasons, and in order to preserve focus and interpretability, we restrict our numerical investigation to the gamma case, which already captures the essential statistical behavior observed in group-level averaging phenomena. Alternative distributions may be explored in future work or used to refine specific empirical applications.

%%%%%%%%%%%%%%%%%%%%%%%%%%%%%%%%%%%%%%%%%%%%%%%%%%%%%%%%%%%%%%%%
\begin{figure}[t]
\begin{center}
\vspace{6mm}
\includegraphics[width=0.5\textwidth,angle=0]{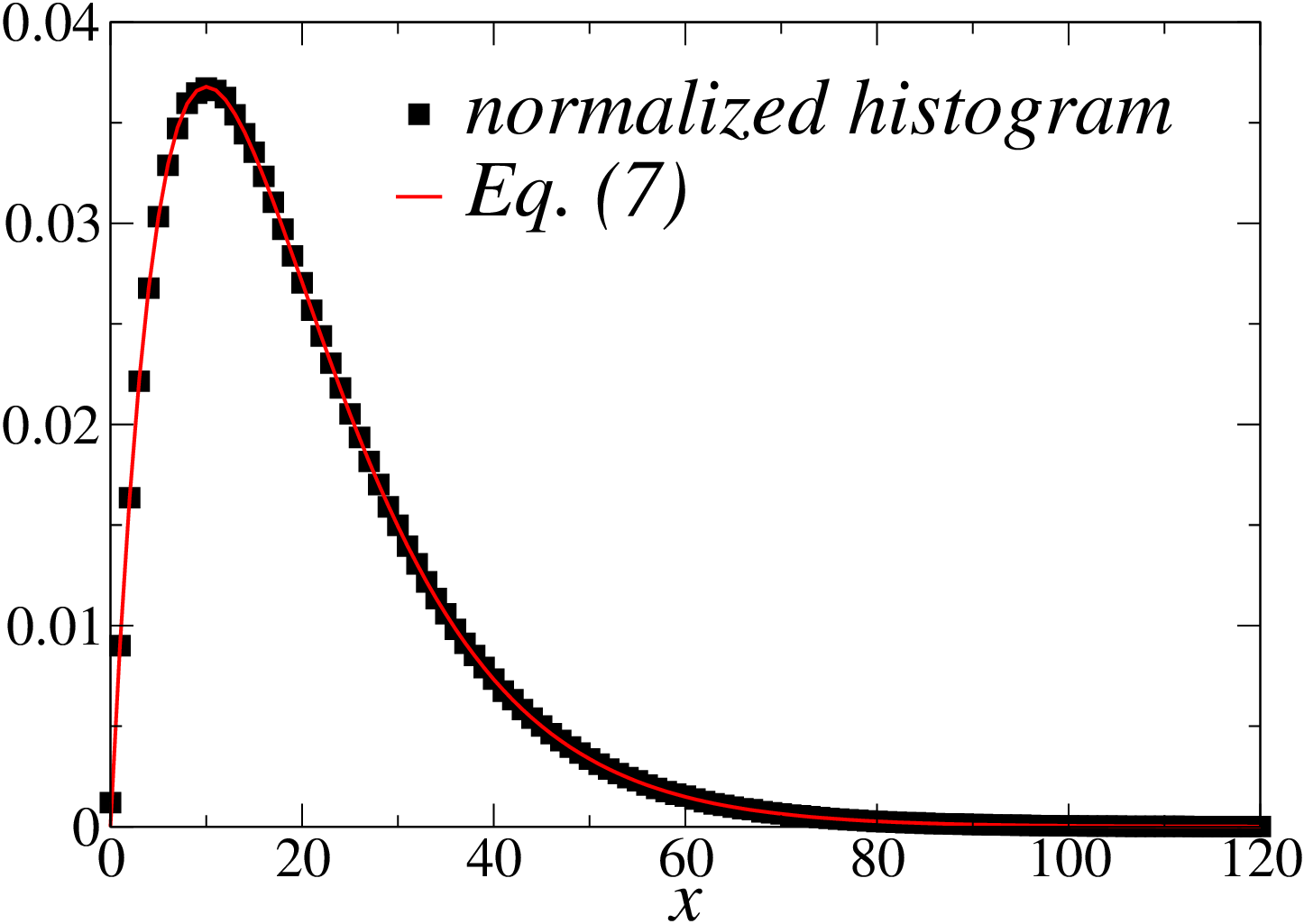}
\end{center}
\caption{Normalized histogram (squares), constructed by dividing the number of occurrences in each bin by the total number of generated values, for gamma-distributed random numbers $x$ obtained numerically as described in the text. For comparison, we plot the analytical gamma distribution given by Eq. \eqref{eq7} (solid line). The parameters are $k=2.0$ and $\theta=10.0$, and $10^{7}$ random values were generated.}
\label{fig1}
\end{figure}
%%%%%%%%%%%%%%%%%%%%%%%%%%%%%%%%%%%%%%%%%%%%%%%%%%%%%%%%%%%%%%%%

To test the analytical predictions, we performed numerical simulations of the model. For the numerical results, the random variables following a gamma distribution were generated using the Marsaglia-Tsang algorithm \cite{mt}, which provides an efficient and widely adopted method for sampling gamma-distributed values. For all numerical results, we considered $k=2.0$ and $\theta=10.0$. Given a group of size $N$, we generated the consumption $x_i$ of each individual as an independent random variable. Subsequently, we calculated the canonical quantities of interest, namely the partition function $Z_{\alpha}$, the mean $\langle x\rangle_{\alpha}$, the variance $\sigma^{2}_{\alpha}$ and the entropy $S_{\alpha}$ of the system, as defined below,
\begin{eqnarray} \label{eq12}
Z_{\alpha} & = & \sum_{i}e^{-\alpha\,x_i} \\ \label{eq13}
\langle x\rangle_{\alpha} & = & \frac{\sum_{i}x_i\,e^{-\alpha\,x_i}}{Z_{\alpha}} \\ \label{eq14}
\sigma^{2}_{\alpha} & = & \frac{\sum_{i}x_i^{2}\,e^{-\alpha\,x_i}}{Z_{\alpha}} - \left(\frac{\sum_{i}x_i\,e^{-\alpha\,x_i}}{Z_{\alpha}}\right)^{2} \\ \label{eq15}
S_{\alpha} & = & \alpha\,\langle x\rangle_{\alpha} + \ln Z_{\alpha}
\end{eqnarray}  
\noindent
The summations are performed from $i=1$ to $i=N$, where $N$ is the group size.

First of all, we present in Fig. \ref{fig1} the numerical histogram of random numbers generated from a gamma distribution with parameters $k=2.0$ and $\theta=10.0$, normalized by the total number of generated samples ($10^{7}$ values). For comparison, we also show the corresponding analytical density function $\rho(x)$ given by Eq. \eqref{eq7}, demonstrating the excellent performance of the Marsaglia-Tsang algorithm.

\begin{figure}[t]
\begin{center}
\vspace{6mm}
\includegraphics[width=0.35\textwidth,angle=0]{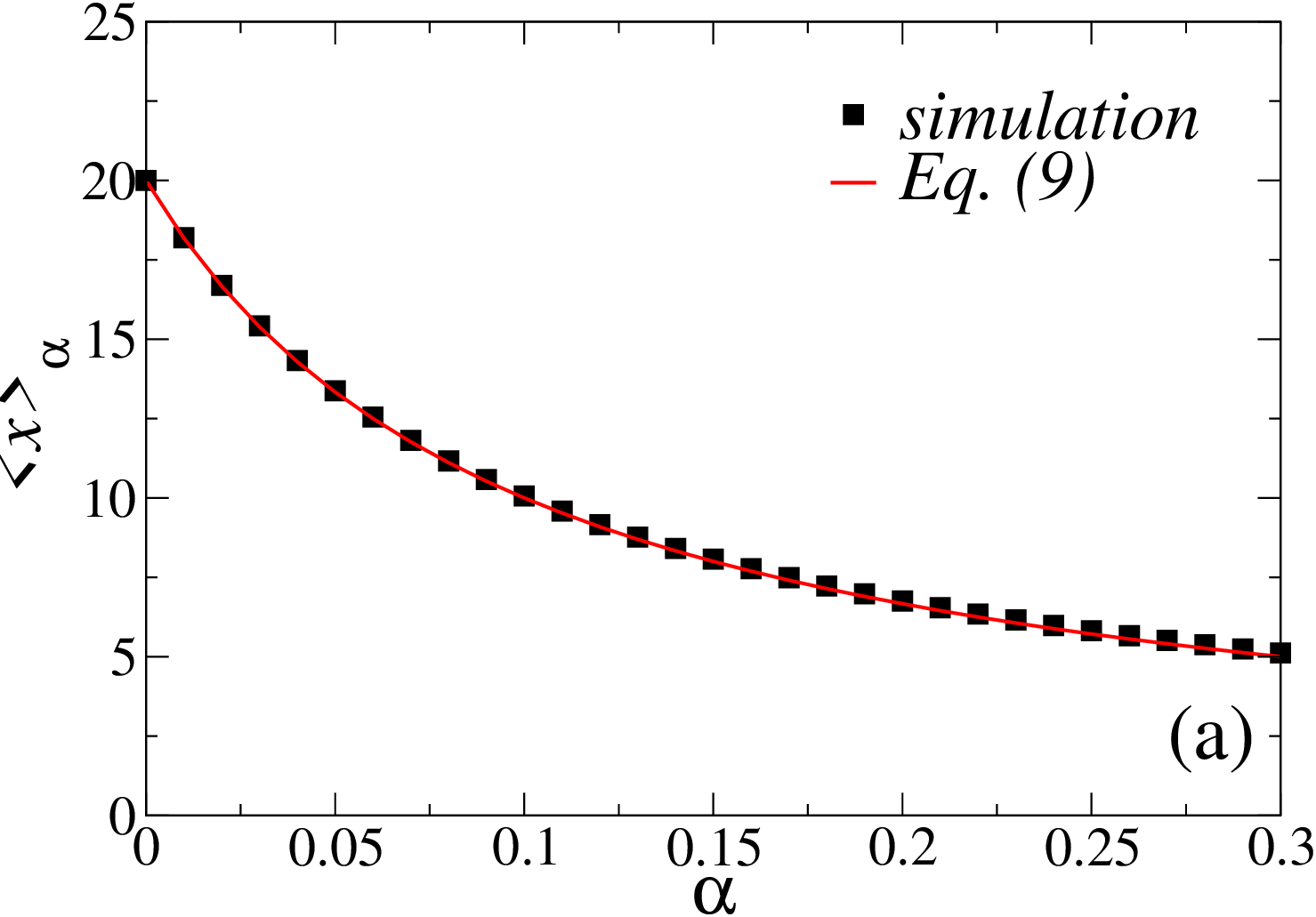}
\hspace{0.3cm}
\includegraphics[width=0.35\textwidth,angle=0]{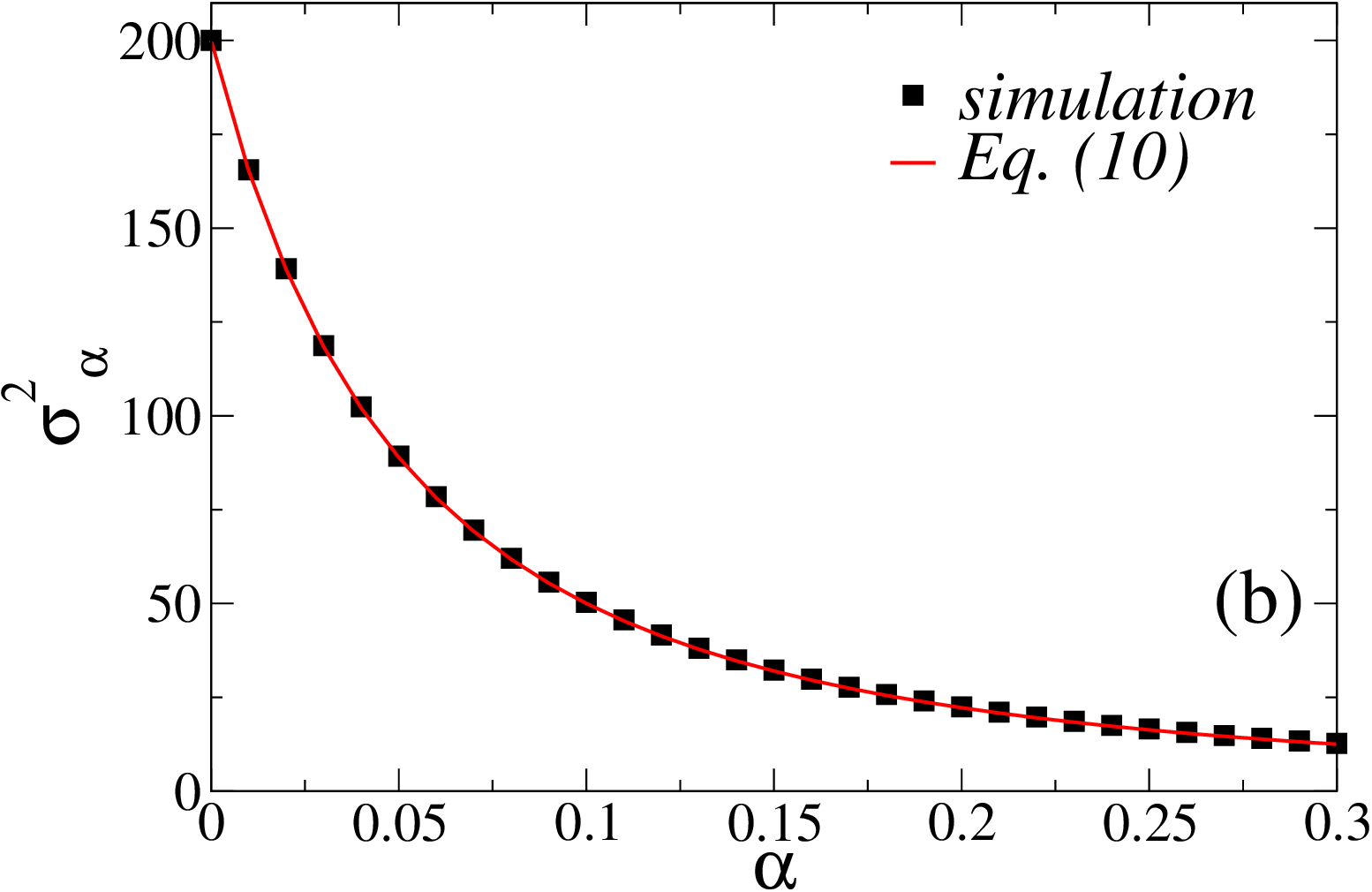}
\\
\vspace{1.0cm}
\includegraphics[width=0.35\textwidth,angle=0]{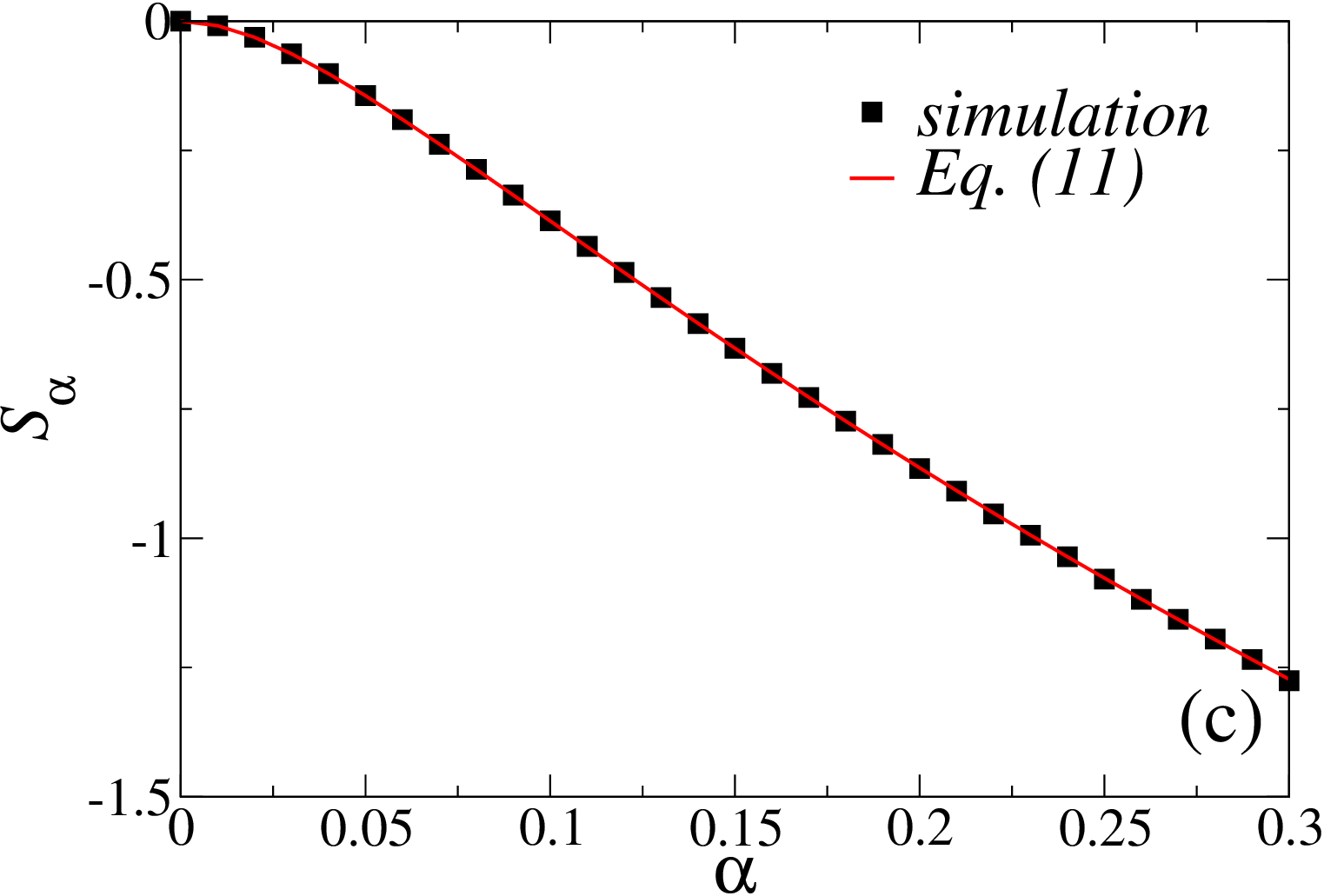}
\end{center}
\caption{Comparison between analytical (solid lines) and numerical (squares) results for three key thermodynamic quantities as functions of the penalization parameter $\alpha$: (a) the canonical average consumption $\langle x\rangle_{\alpha}$, (b) the canonical variance $\sigma^{2}_{\alpha}$, and (c) the canonical entropy $S_{\alpha}$. Numerical results were obtained from simulations using $10^{6}$ samples per value of $\alpha$, with error bars smaller than the symbol sizes. The excellent agreement between analytical predictions and numerical estimates confirms the validity of the theoretical framework and the robustness of the computational implementation. The group size is $N=10$.}
\label{fig2}
\end{figure}

We show in Fig. \ref{fig2} the behavior of three key quantities as functions of the penalization parameter $\alpha$. Panel (a) displays the canonical average consumption $\langle x\rangle_{\alpha}$, panel (b) shows the canonical variance $\sigma^{2}_{\alpha}$, and panel (c) presents the canonical entropy $S_{\alpha}$. In each panel, analytical predictions (solid lines), obtained from Eqs. \eqref{eq9} - \eqref{eq11}, are compared with numerical estimates (squares), demonstrating excellent agreement across the entire range of $\alpha$. As expected, the average consumption decreases monotonically with increasing $\alpha$, reflecting the stronger penalization of high expenditures. The variance exhibits a similar decay, indicating that individual behaviors become more homogeneous. The entropy decreases monotonically as well, assuming negative values for all $\alpha>0$. This reflects the progressive restriction of accessible microstates as the penalization favors lower individual expenditures, effectively reducing the system's configurational freedom. In the statistical mechanics analogy, this behavior corresponds to the ordering process observed in low-temperature phases, where the system becomes increasingly concentrated in a small set of low-energy states. The negative values of the entropy, while unusual in classical thermodynamics, are a consequence of the present definition relative to a unit measure; they indicate that the weighted distribution is more concentrated than the original unweighted gamma distribution, emphasizing the emergence of order from an initially disordered system.

%%%%%%%%%%%%%%%%%%%%%%%%%%%%%%%%%%%%%%%%%%%%%%%%%%%%%%%%%%%%%%%%
%%%%%%%%%%%%%%%%%%%%%%%%%%%%%%%%%%%%%%%%%%%%%%%%%%%%%%%%%%%%%%%%

We also performed simulations of the model for a fixed penalization parameter $\alpha$ and various group sizes $N$, in order to estimate the relative deviation $\delta_{N}$ of the group-averaged expenditure as a function of $N$. Specifically, we compute the variance of the canonical average over several independent realizations of groups. The relative deviation of the group-averaged consumption is defined as
\begin{equation} \label{eq16}
\delta_{N} = \frac{\sigma(\bar{x}_{N})}{\langle \bar{x}_{N}\rangle} ~,
\end{equation}
where $\bar{x}_{N}=(1/N)\sum_{i}\,x_i$ is the average consumption for a group size $N$, $\langle \bar{x}_{N}\rangle$ denotes its expected value, and $\sigma(\bar{x}_{N})$ is its standard deviation. All calculated quantities were averaged over $10^{6}$ independent simulations for each value of $N$.

%On the distinction between individual and collective fluctuations.
It is important to distinguish between two types of statistical fluctuations considered in this work. When studying the relative deviation $\delta_{N}$ of the group-averaged expenditure as a function of the group size $N$, each realization corresponds to a group of $N$ individuals with expenditures $\{x_i\}$ sampled independently from a gamma distribution. The group average is computed using canonical weights $e^{-\alpha\,x_i}$, reflecting the influence of the external parameter $\alpha$. The quantity $\delta_{N}$ is then defined as the relative deviation of these weighted averages over many realizations. This approach is consistent with a canonical ensemble of weakly interacting agents, and we expect to observe a decay of the form $\delta_{N}\sim N^{-a}$, with some exponent $a$, reflecting the expected suppression of collective fluctuations. On the other hand, when computing the canonical variance of the individual variable $x$, the appropriate procedure is to apply the Boltzmann weights directly to each $x_i$ over a large sample, without grouping. Mixing these two approaches may lead to incorrect conclusions about the behavior of statistical moments under canonical weighting.

\begin{figure}[t]
\begin{center}
\vspace{6mm}
\includegraphics[width=0.5\textwidth,angle=0]{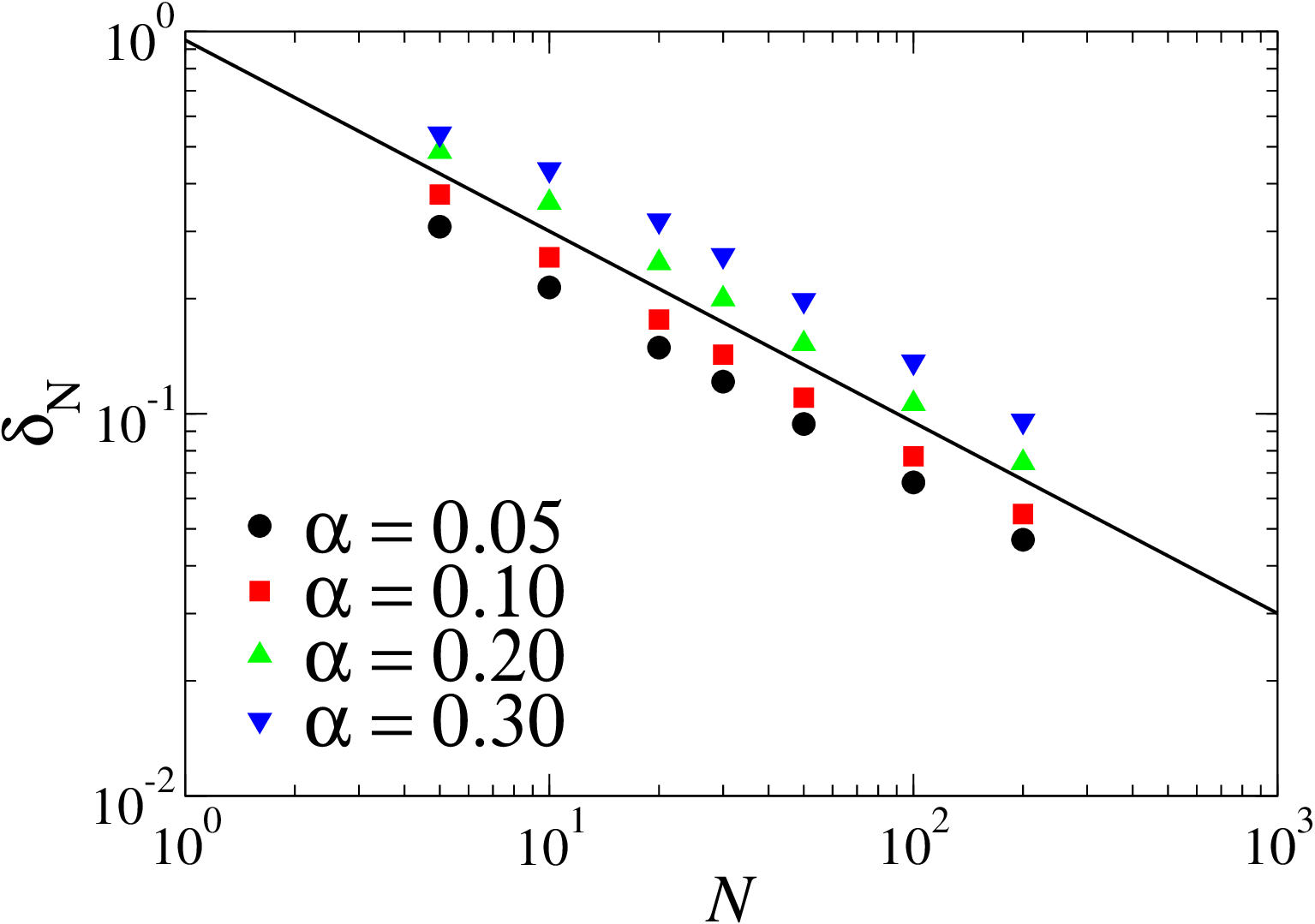}
\end{center}
\caption{Relative deviation $\delta_{N}$, given by Eq. \eqref{eq16}, vs group size $N$, for typical values of $\alpha$, plotted in the log-log scale. As a guide to the eye, a straight line with slope $-0.5$ is shown. Each point represents an average over $10^{6}$ independent simulations.}
\label{fig3}
\end{figure}

In Fig. \ref{fig3} we plot $\delta_{N}$ vs $N$ for typical values of $\alpha$. As it can be seen, our numerical results confirm that, regardless of the value of $\alpha$, the relative deviation consistently scales as $\delta_{N}\sim N^{-a}$, with the exponent $a$ very close to $1/2$. This finding is in agreement with the central limit theorem and indicates that, even under canonical weighting by the Boltzmann factor $e^{-\alpha\,x}$, individual expenditures remain effectively independent. The system thus retains the statistical extensivity of an ideal gas, where aggregate fluctuations are suppressed as the number of individuals increases \cite{huang}. Thus, as $N$ increases, deviations shrink and fairness in the bill improves.

%%%%%%%%%%%%%%%%%%%%%%%%%%%%%%%%%%%%%%%%%%%%%%%%%%%%%%%%%%%%%%%%
%%%%%%%%%%%%%%%%%%%%%%%%%%%%%%%%%%%%%%%%%%%%%%%%%%%%%%%%%%%%%%%%

As a numerical example, let us consider the same parameters for the gamma distribution adopted throughout the text, i.e., $k=2.0$ and $\theta=10.0$. First, for the unpenalized case ($\alpha=0.0$), the average consumption is given by Eq. \eqref{eq9} with $\alpha=0.0$, yielding $\langle x\rangle_{\alpha=0.0} = k\,\theta= \$\,20.00$, with standard deviation given by Eq. \eqref{eq10}, namely $\sigma_{\alpha=0.0}=\sqrt{k\,\theta^{2}}=\$\,14.14$. Introducing a moderate penalization ($\alpha=0.05$) reduces the average to $\$\,13.33$ and the standard deviation to $\$\,9.43$. {Considering a stronger penalization such as $\alpha=0.30$ further decreases the average to $\$\,5.00$ with a standard deviation of only $\$\,3.54$. For a group of size $N=10$, for example, the fluctuation of the group-averaged bill, given by $\sigma_{\bar{x}_N}=\sqrt{\sigma^{2}_{\alpha}/N}$, reduces accordingly from $\$\,4.47$ in the unpenalized case ($\alpha=0.0$) to $\$\,2.98$ at $\alpha=0.05$ and $\$\,1.12$ at $\alpha=0.30$. This illustrates how penalization leads to both lower average expenditures and increased collective homogeneity. Moreover, as previously discussed, the fluctuation of the group-averaged bill decreases with group size, reinforcing the fairness of each individual paying the group's average consumption. A summary of these results is presented in Table \ref{Tab2}, where we also include the group mean standard deviation for $N=30$ to enable comparison with the results for group size $N=10$. It can be seen that increasing the group size further reduces fluctuations in the group-averaged bill, thereby reinforcing the statistical fairness of equal cost-sharing.

%%%%%%%%%%%%%%%%%%%%%%%%%%%%%%%%%%%%%%%%%%%%%%%%%%%%%%%%%%%%%%%%%%%%%%%%
\begin{table*}[tbp]
\begin{center}
\vspace{0.1cm}
\renewcommand\arraystretch{0.9} 
\begin{tabular}{|c|c|c|c|c|}
\hline
$\alpha$ & $\langle x\rangle_{\alpha}$ & $\sigma_{\alpha}$ & $\sigma_{\bar{x}_{10},\alpha}$ & $\sigma_{\bar{x}_{30},\alpha}$ \\ \hline
$0.00$ & $\$\,20.00$ & $\$\,14.14$ & $\$\,4.47$ & $\$\,2.58$ \\
$0.05$ & $\$\,13.33$ & $\$\,9.43$  & $\$\,2.98$ & $\$\,1.72$ \\
$0.10$ & $\$\,10.00$ & $\$\,7.07$  & $\$\,2.24$ & $\$\,1.29$ \\
$0.20$ & $\$\,6.67$ & $\$\,4.71$  & $\$\,1.49$ & $\$\,0.86$ \\
$0.30$ & $\$\,5.00$ & $\$\,3.54$  & $\$\,1.12$ & $\$\,0.65$ \\
%$0.15$ & $\$\,8.00$  &  $\$\,5.66$ & $\$\,1.79$ & $\$\,1.03$ \\
\hline
\end{tabular}%
\end{center}
\caption{Comparison of individual average consumption $\langle x\rangle_{\alpha}$, individual standard deviation $\sigma_{\alpha}$, and group mean standard deviation $\sigma_{\bar{x}_{N},\alpha}$ for different values of the penalization parameter $\alpha$, considering a gamma distribution with parameters $k=2.0$ and $\theta=10.0$. Results are shown for two group sizes, $N=10$ and $N=30$, illustrating how increasing group size further reduces fluctuations in the group-averaged bill, thereby reinforcing the statistical fairness of equal cost-sharing. Moreover, increasing $\alpha$ reduces both the mean expenditure and its variability, highlighting the ``ordering'' effect of penalization.}
\label{Tab2}
\end{table*}
%%%%%%%%%%%%%%%%%%%%%%%%%%%%%%%%%%%%%%%%%%%%%%%%%%%%%%%%%%%%%%%%%%%%%%%%

These results are corroborated by previous researches into social behavior that consider the bill-splitting scenario. In fact, many studies point to the fact that in groups, specially large ones, food and drink consumers tend to flock to particular categories of items (even if choosing different items themselves)~\cite{ellison}, which can lead to discrepancies that average out when the whole bill is divided~\cite{woolley}. This was what was observed in studies made only with receipts, but in more careful probing using the examination of who ordered which item first, it was further observed that the leading orders tend to influence the other consumers in the group in their respective choices~\cite{schamel,leekimkwak}. In fact, both the social experience and the social ties between the consumers seem to be enhanced by the bill splitting~\cite{woolley}.

On the other hand, it was also shown very recently that the expected happiness of paying a bill in full for a cherished person is even more rewarding~\cite{smithbarton}. Nevertheless, when grouped with someone with weaker social ties, the splitting of the bill can still be the preferred option. Furthermore, a study by Y. Shani~\cite{shani} corroborates the implications proposed by Gneezy \textit{et al.}~\cite{gneezy}, but in a controlled situation where people are questioned solely about buying chocolate candy bars, which does not reproduce the informality and spontaneity of a dining environment. Still, one must take into account that what the results here show is a tendency of the discrepancies between different consumptions to even out when splitting a bill equally - the sharing experience and the influence of previous orders~\cite{ellison,woolley,schamel,leekimkwak} are maybe natural social causes of that result, which in turn are modeled here by a thermodynamical system. Of course, this differs considerably from both the works of Gneezy {\it et al.}~\cite{gneezy} and Shani~\cite{shani} and from market econophysical models, where the change from a social to a market-influenced setting leads to a competitive behavior for which our model is not intended~\cite{reeson}.

%%%%%%%%%%%%%%%%%

\section{Discussion and conclusions}

The almost natural perception of fairness in equal cost-sharing in social settings emerges as a natural consequence of statistical averaging, analogous to energy distributions in thermodynamic ensembles. This perspective not only demystifies a common social behavior but also illustrates the power of statistical physics in explaining emergent fairness in economic and social systems.

To sum up the previous sections, in this work we considered a model based on an analogy with the ideal gas in the canonical ensemble. In this framework, the consumption $x_i$ of a given individual is analogous to the kinetic energy of a gas particle $E_i$, and the average consumption corresponds to the average energy per particle. A Boltzmann-like factor is introduced, namely $e^{-\alpha\,x_i}$, where the parameter $\alpha$ can be interpreted as a penalization factor. Taking into account a functional entropy form, and the maximum relative entropy method, the parameter $\alpha$ was identified as the Lagrange multiplier used in the maximization procedure of the functional over the probability $P(x_i)$, subject to normalization and the constraint given by the average consumption. The parameter $\alpha$ plays a role which is analogous to the inverse temperature $\beta=1/T$ in statistical mechanics. At low $\alpha$ (high temperature), individual expenditures are weakly constrained and the system exhibits large fluctuations, akin to a high-temperature gas where particles have a broad energy distribution. Conversely, at high $\alpha$ (low temperature), the Boltzmann weighting strongly penalizes large expenditures, inducing a more concentrated, ``ordered'' distribution of behavior, analogous to how particles in a low-temperature gas predominantly occupy lower-energy states. We considered the gamma distribution for the density of states, where an state in the model is given by the expenditures of the $N$ individuals in the group, i.e., $\{x\}=\{x_1,x_2, ..., x_N\}$.

We derived analytical results for the canonical partition function $Z_{\alpha}$, the average consumption $\langle x\rangle_{\alpha}$ and its variance $\sigma^{2}_{\alpha}$, with all quantities calculated considering the Boltzmann-like factor $e^{-\alpha\,x_i}$. The calculations were performed using the gamma distribution, as it appears to realistically represent the problem of bar consumption. An important result is that the average consumption  $\langle x\rangle_{\alpha}$ is inversely proportional to $\alpha$, highlighting the key role of the penalization parameter: for high $\alpha$, individuals in the group tend to restrict their expenditures to lower values, leading to a decrease in the average consumption as everyone converges towards similar spending levels. This is analogous to the behavior of an ideal gas at low temperature (high $\beta=1/T$), where particles predominantly occupy low-energy states. Conversely, for low $\alpha$ (high temperature), individuals tend not to constrain their expenditures, resulting in higher average consumption and greater variability, similar to how particles in a high-temperature gas populate a broad range of energies.

%We derived analytical results for the canonical partition function $Z_{\alpha}$, the average consumption $\langle x\rangle_{\alpha}$ and its variance $\sigma^{2}_{\alpha}$, with all quantities calculated considering the Boltzmann-like factor $e^{-\alpha\,x_i}$. The calculations were performed using a gamma distribution, as it appears to realistically represent the problem of bar consumption. An important result is that the average consumption $\langle x\rangle_{\alpha}$ is inversely proportional to $\alpha$, highlighting the key role of the penalization parameter: for high $\alpha$, individuals in the group tend to ``align'' their behaviors with one another, leading to a decrease in the average consumption, as everyone tends to consume similar items. This is analogous to the alignment observed in the low-temperature (high $\beta=1/T$) phase of the Ising model \cite{huang}. Conversely, for low $\alpha$, individuals tend not to care about the judgment of others in the group, and the tendency to consume expensive items increases, resulting in a higher mean consumption. This resembles the Ising model at high temperature, where spins do not tend to align with neighbor spins.

Numerical simulations for various values of the penalization parameter $\alpha$ confirm that the relative deviation of the group-averaged bill $\delta_{N}$ of the canonical average scales with the group size $N$ as $\delta\sim N^{-a}$, with the exponent $a$ being very close to $1/2$. This indicates that statistical independence among individual expenditures is preserved even under canonical weighting, consistent with the law of large numbers \cite{reif}. Such behavior highlights the robustness of the collective average against individual variability, further supporting the analogy with extensive thermodynamic systems \cite{huang}. In fact, the social aspect of a bar bill interaction, e.g. noticing if one's consumption differs too much from the others may hinder this ones next orders, may contribute to regulate the system as a whole and keep it inside the low temperature regime (i.e. high $\alpha$) of the canonical ensemble.

Thus, even though individuals make autonomous, diverse choices about what to consume, the collective outcome displays regularity, predictability, and a sense of equity. This is not imposed externally, but emerges from the interplay of statistical laws, weak social correlations, and ensemble behavior. This regularity is not accidental: it is a resemblance to complex systems where microscopic randomness coupled to self-organizing (in this case, social) mechanisms, such as a sense of guilt in spending much more than others, gives rise to macroscopic predictability. Thus, what seems like a simple social custom is in fact rooted in the same principles that govern energy distributions in matter.

In fact, agents operate based on local rules or preferences, yet the system as a whole exhibits coherence and structure \cite{pmco}. The emergence of fairness from randomness mirrors other examples in nature and society, such as traffic flows, consensus formation, and opinion dynamics. In our model, equal division of the total cost becomes a reliable and robust rule, not because everyone consumes the same, but because the aggregate behavior smooths out individual variation. This insight reinforces the value of statistical physics as a conceptual lens for understanding not just physical matter, but the organized patterns of social and economic life. The simple act of splitting a restaurant bill among friends can be reinterpreted as a thermodynamic system in statistical equilibrium under social constraints. What seems like an informal social convention -- dividing the total cost equally -- emerges naturally from simple statistical principles: mutual compensation, suppression of fluctuations, and ensemble averaging.

\section*{Acknowledgments}

N. Crokidakis acknowledges partial financial support from the Brazilian scientific funding agency Conselho Nacional de Desenvolvimento Científico e Tecnológico, Brazil (CNPq, Grant 308643/2023-2). L. Sigaud acknowledges partial financial support from the Brazilian scientific funding institute INCT-FNA. We are indebted to an anonymous referee who drew our attention to the application of the maximum relative entropy method to our model.

\bibliographystyle{elsarticle-num-names}

\end{document}